\documentclass[sigconf]{acmart}
\usepackage[switch]{lineno}
\usepackage[utf8]{inputenc}
\usepackage{tabularray}
\usepackage{booktabs}
\usepackage{multirow}
\usepackage{makecell}
\usepackage{float}
\usepackage{hyperref}
\usepackage{enumitem}

\AtBeginDocument{%
  }

\copyrightyear{2026}
\acmYear{2026}
\setcopyright{cc}
\setcctype{by}
\acmConference[WSDM '26]{Proceedings of the Nineteenth ACM International Conference on Web Search and Data Mining}{February 22--26, 2026}{Boise, ID, USA}
\acmBooktitle{Proceedings of the Nineteenth ACM International Conference on Web Search and Data Mining (WSDM '26), February 22--26, 2026, Boise, ID, USA}
\acmPrice{}
\acmDOI{10.1145/3773966.3778012}
\acmISBN{979-8-4007-2292-9/2026/02}
\begin{document}

\title{Time-Aware Adaptive Side Information Fusion for Sequential Recommendation}

\author{Jie Luo}
\orcid{0009-0003-5125-0560}
\affiliation{%
  \institution{University of Science and Technology of China}
  \city{Hefei}
  \country{China}
}
\email{luojie2000@mail.ustc.edu.cn}

\author{Wenyu Zhang}
\orcid{0009-0001-1457-1707}
\affiliation{%
  \institution{University of Science and Technology of China}
  \city{Hefei}
  \country{China}
}
\email{wenyuz@mail.ustc.edu.cn}

\author{Xinming Zhang}
\authornote{Corresponding authors.}
\orcid{0000-0002-8136-6834}
\affiliation{%
  \institution{University of Science and Technology of China}
  \city{Hefei}
  \country{China}
}
\email{xinming@ustc.edu.cn}

\author{Yuan Fang}
\authornotemark[1]
\orcid{0000-0002-4265-5289}
\affiliation{%
  \institution{Singapore Management University Singapore}
  \country{Singapore}
}
\email{yfang@smu.edu.sg}

\renewcommand{\shortauthors}{Jie Luo, Wenyu Zhang, Xinming Zhang, and YuanFang}

\begin{abstract}
Incorporating item-side information, such as category and brand, into sequential recommendation is a well-established and effective approach for improving performance. However, despite significant advancements, current models are generally limited by three key challenges: they often overlook the fine-grained temporal dynamics inherent in timestamps, exhibit vulnerability to noise in user interaction sequences, and rely on computationally expensive fusion architectures. To systematically address these challenges, we propose the \textbf{T}ime-Aware \textbf{A}daptive \textbf{S}ide \textbf{I}nformation \textbf{F}usion (\textbf{TASIF}) framework. TASIF integrates three synergistic components: (1) a simple, plug-and-play time span partitioning mechanism to capture global temporal patterns; (2) an adaptive frequency filter that leverages a learnable gate to denoise feature sequences adaptively, thereby providing higher-quality inputs for subsequent fusion modules; and (3) an efficient adaptive side information fusion layer, this layer employs a "guide-not-mix" architecture, where attributes guide the attention mechanism without being mixed into the content-representing item embeddings, ensuring deep interaction while ensuring computational efficiency. Extensive experiments on four public datasets demonstrate that TASIF significantly outperforms state-of-the-art baselines while maintaining excellent efficiency in training. Our source code is available at \url{https://github.com/jluo00/TASIF}.
\end{abstract}

\ccsdesc[500]{Information systems~Recommender systems}

\keywords{Sequential Recommendation, Side Information Fusion, Temporal Information}


\maketitle

\section{INTRODUCTION}\label{Sec: intro}
\begin{figure}[t]
  \centering
   \includegraphics[width=\linewidth]{images/fig1.pdf}
   \caption{Illustration of the Time Span Partitioning (TSP) mechanism. We partition the global timeline into pre-defined spans (e.g., month or quarter) and assign a corresponding time token to each user interaction based on its timestamp.}
   \Description{}
   \label{fig1: TSP}
\end{figure}

\begin{figure}[t]
  \centering
   \includegraphics[width=\linewidth]{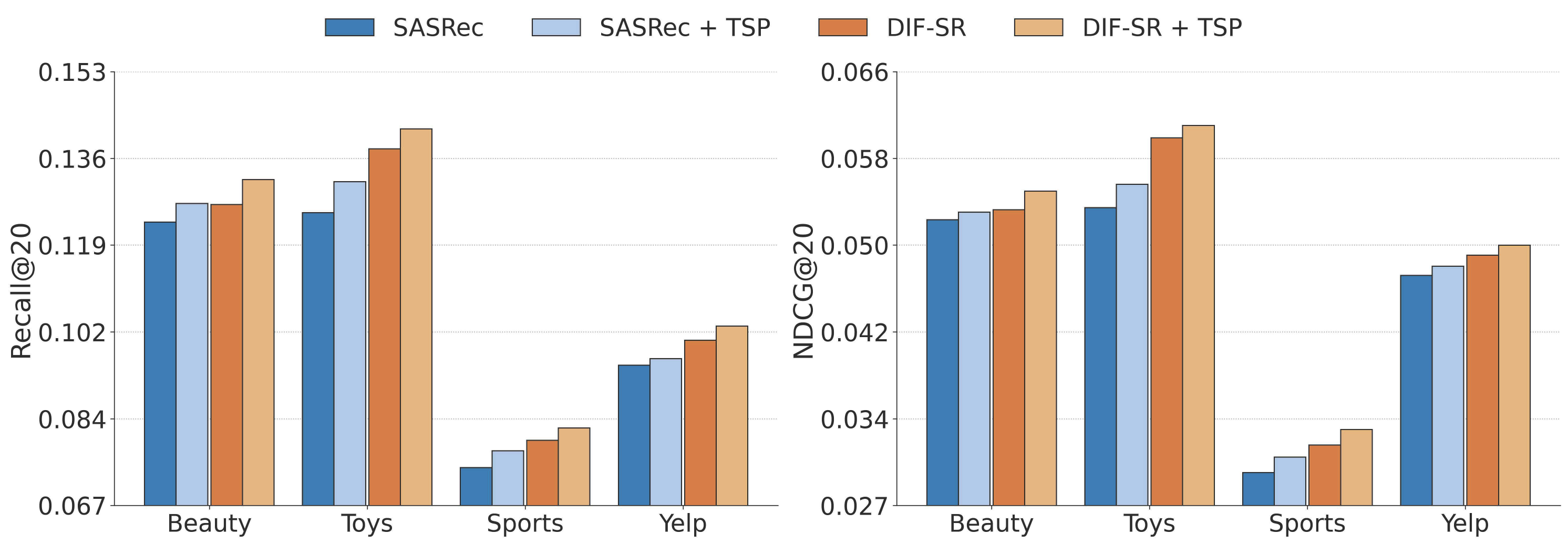}
   \caption{Our plug-and-play TSP module effectively integrates with and consistently boosts the performance of mainstream sequential recommendation methods like SASRec and DIF-SR on benchmark datasets.}
   \Description{}
   \label{fig2: ablation_tsp}
\end{figure}

Sequential recommendation (SR) is a fundamental task in recommendation systems, aiming to model the dynamics of user behavior to capture evolving interests and enhance recommendation performance \cite{SR_survey_1:conf/ijcai/WangHWCSO19,SR_survey_2:journals/csur/WangCWSOL22,SR_survey_3:journals/corr/abs-2412-12770}. Driven by the rapid advances in deep learning, modeling techniques for SR have undergone a significant evolution. This progression ranges from early methods based on Recurrent Neural Networks (RNNs) for capturing sequence dependencies \cite{GRU4Rec:journals/corr/HidasiKBT15}, to the use of Convolutional Neural Networks (CNNs) for extracting local interaction patterns \cite{Caser:conf/wsdm/TangW18}, and more recently, to the adoption of Transformer architectures \cite{Transformer:conf/nips/VaswaniSPUJGKP17} that leverage attention mechanisms to model global dependencies within sequences \cite{SASRec:conf/icdm/KangM18, BERT4Rec:conf/cikm/SunLWPLOJ19, TiSASRec:conf/wsdm/LiWM20}.

While sequential recommendation has witnessed substantial advancements, existing methods remain constrained by their heavy reliance on item ID modeling while underutilizing rich side information (e.g., categories, brands) in user behavior sequences. To address this limitation, researchers have explored side information fusion mechanisms. NOVA~\cite{NOVA:journals/corr/abs-2103-03578} pioneered a ``non-invasive'' attention, using attribute information only for attention weight computation while preserving ID embeddings as independent values to prevent information contamination. Building on this, DIF-SR~\cite{DIF-SR:conf/sigir/XieZK22} achieved personalized learning for side information through decoupled attention matrices. MSSR~\cite{MSSR:conf/wsdm/LinLPP0LH024} then introduced additional cross-sequence attention and self-supervised alignment tasks to optimize representations. More recently, state-of-the-art models like DIFF~\cite{DIFF:conf/sigir/Kim00BL25} have further advanced the field by proposing a dual-fusion architecture that combines intermediate and early fusion.

\begin{figure*}[ht]
  \centering
   \includegraphics[width=0.95\linewidth]{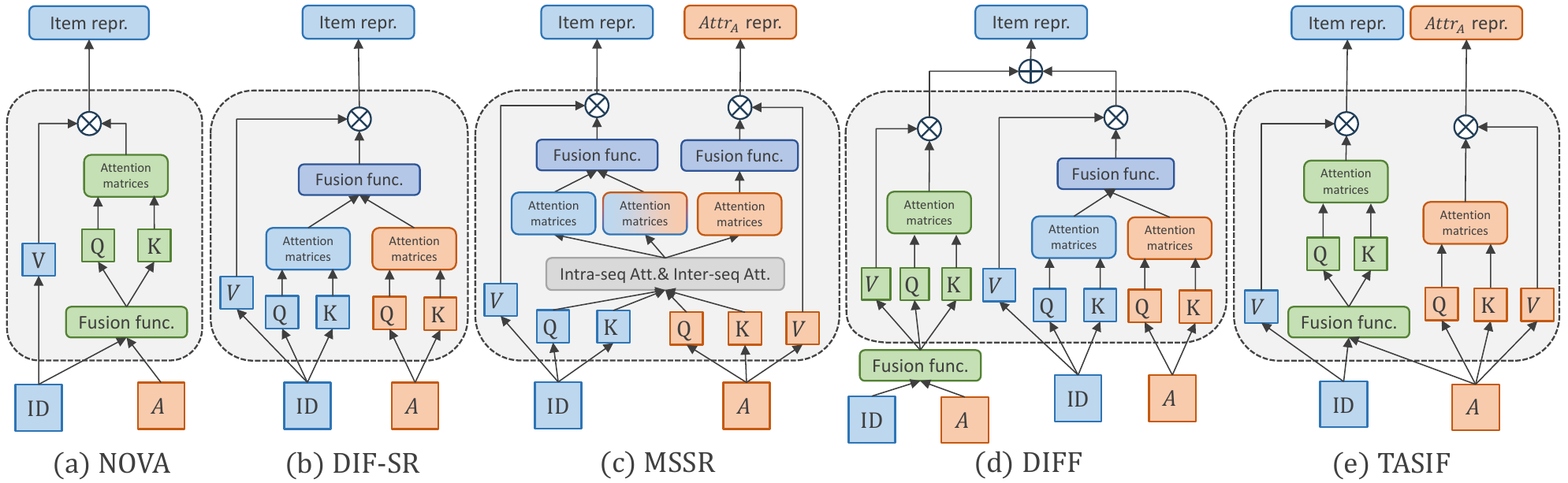}
   \caption{Comparison of different side information fusion methods. NOVA maintains value independence while fusing item representations; DIF-SR decouples attention computation and fused attention matrices; MSSR introduces a multi-sequence cross-attention mechanism; DIFF adopts a dual-fusion architecture; our proposed TASIF employs a lightweight design that decouples item-attribute updates and uses attributes to guide item attention.}
   \Description{}
   \label{fig3:dif}
\end{figure*}

However, in their pursuit of performance, these advanced side-information fusion models exhibit increasingly complex structural designs, which in turn exposes several fundamental limitations:
\textbf{(1) A systemic neglect of temporal dynamics.} Most advanced fusion models largely overlook the effective modeling of rich dynamic patterns embedded in precise timestamps. Although dedicated time-aware models exist, they either focus on fine-grained inter-item intervals (e.g., TiSASRec~\cite{TiSASRec:conf/wsdm/LiWM20}) or devise complex time-aware attention structures (e.g., TimelyRec~\cite{TimelyRec:conf/www/ChoHKY21}), making their integration into already-bloated fusion frameworks impractical. Meanwhile, treating timestamps as just another piece of side information (e.g., DLFS-Rec~\cite{DLFS-Rec:conf/recsys/LiuD0P023}) fails to capture their unique global contextual information;
\textbf{(2) The sub-optimal handling of noise in feature sequences.} Due to the inherent noise in feature sequences (e.g., irrelevant interactions), fusing additional side information often exacerbates the problem of noise contamination. While some models like DIFF~\cite{DIFF:conf/sigir/Kim00BL25} have begun to employ frequency-domain filters, they rely on a static strategy, such as uniformly suppressing high frequencies. This "one-size-fits-all" approach is overly rigid, unable to distinguish between noise and valuable signals like short-term interest bursts that may coexist in high-frequency components;
\textbf{(3) The efficiency bottleneck of the fusion mechanism itself.} To maximize the utility of side information, some models resort to computationally intensive cross-sequence attention (e.g., MSSR~\cite{MSSR:conf/wsdm/LinLPP0LH024} shown in Fig.~\ref{fig3:dif}(c)), while others adopt parallel dual-fusion pathways (e.g., DIFF~\cite{DIFF:conf/sigir/Kim00BL25} shown in Fig.~\ref{fig3:dif}(d)). Although these designs improve performance, they introduce significant computational overhead, limiting their potential for real-world applications.

To systematically address the limitations mentioned above, we propose a novel framework, named \textbf{T}ime-Aware \textbf{A}daptive \textbf{S}ide \textbf{I}nformation \textbf{F}usion (\textbf{TASIF}), that integrates time awareness, adaptive denoising, and an efficient fusion mechanism.

For \textit{time awareness,} inspired by Wu \emph{et al.}~\cite{SimpleDyg:conf/www/WuFL24}, we design a novel, simple, and plug-and-play Time Span Partitioning (TSP) mechanism as shown in Fig.~\ref{fig1: TSP}. TSP discretizes the global timeline into uniform spans, providing a shared temporal anchor to align different user sequences and capture periodic behaviors. As illustrated in Fig.~\ref{fig2: ablation_tsp}, TSP, as a standalone and universal module, consistently boosts the performance of various baselines, proving its wide applicability.

For \textit{adaptive denoising,} to overcome the limitations of existing
filtering strategies, we propose a Adaptive Frequency Filter (AFF) layer. It adopts learnable filters for refined frequency-domain analysis and replaces the conventional residual connection with a learnable gate. This gate enables the adaptive fusion of original and denoised features, allowing the model to dynamically adjust denoising intensity based on data characteristics and thereby provide higher-quality inputs for downstream fusion modules.

For \textit{efficient fusion,} we design an architecturally simpler and more efficient Adaptive Side Information Fusion (ASIF) layer. Its core lies in decoupling the update paths for the main item and attribute representations. The attribute sequence evolves in an independent pathway, and its refined features act only as a ``guiding signal'' for the main item sequence's attention calculation. This design ensures deep information interaction while drastically reducing computational complexity, achieving a superior balance between performance and efficiency, as shown in Fig.~\ref{fig3:dif}(e).

The main contributions of this work are four-fold:
\begin{itemize}[leftmargin=*]
    \item We propose Time Span Partitioning, a simple, plug-and-play mechanism for SR. By discretizing the timeline into uniform spans, TSP creates a shared temporal anchor that aligns user sequences and captures periodic behaviors.
    \item We propose the Adaptive Frequency Filter, which introduces a learnable gating to dynamically fuse original features with their frequency-denoised counterparts, providing higher-quality input for downstream fusion modules.
    \item We design an Adaptive Side Information Fusion layer whose "guide-not-mix" decoupled architecture achieves deep interaction by fully utilizing side information while maintaining exceptional computational efficiency.
    \item We conducted extensive experiments on four public datasets, which systematically demonstrate that the TASIF model significantly outperforms all baselines. Ablation studies also verify the independent effectiveness of each component.
\end{itemize}

\section{RELATED WORKS}

In this section, we review related work on sequential recommendation and side information fusion for sequential recommendation.

\subsection{Sequential Recommendation}
Sequential recommendation aims to predict users' future interests from their historical interaction sequences. Early methods, such as FPMC \cite{FPMC:conf/www/RendleFS10}, primarily relied on Markov Chains but struggled to model complex dependencies. 
The rise of deep learning brought breakthroughs. GRU4Rec \cite{GRU4Rec:journals/corr/HidasiKBT15} pioneered the use of Recurrent Neural Networks, while Caser \cite{Caser:conf/wsdm/TangW18} and NextItNet \cite{NextItNet:conf/wsdm/YuanKAJ019} applied convolutional and dilated convolutional networks, respectively. The Transformer architecture marked another major advance, with SASRec \cite{SASRec:conf/icdm/KangM18} introducing the self-attention mechanism and BERT4Rec \cite{BERT4Rec:conf/cikm/SunLWPLOJ19} enhancing representations via bidirectional modeling. 

Furthermore, to better leverage temporal information from interactions, time-aware sequential recommendation has been explored. For instance, TiSASRec \cite{TiSASRec:conf/wsdm/LiWM20} uses temporal interval embeddings to capture interest drift, while TiCoSeRec \cite{TiCoSeRec:conf/aaai/DangYGJ0XSL23} handles irregular temporal distributions via contrastive learning with time-homogenizing augmentation. For temporal context encoding, TimelyRec \cite{TimelyRec:conf/www/ChoHKY21} models periodic patterns with multi-scale attention, and MOJITO \cite{Mojito:conf/sigir/TranSSH23} designs a time-disentangled dual-stream architecture to separate long- and short-term preferences, thereby mitigating interest drift.

\subsection{Side Information Fusion for Sequential Recommendation}
In recent years, researchers have explored leveraging side information such as item attributes to enhance sequential recommendation performance. Early studies mainly adopted independent modeling strategies: FDSA \cite{FDSA:conf/ijcai/ZhangZLSXWLZ19} processed item-level and feature-level sequences through separate self-attention modules; S$^3$Rec \cite{S3Rec:conf/cikm/ZhouWZZWZWW20} proposed a self-supervised pre-training framework to learn item-side information associations. However, these independent modeling methods struggle to capture sufficient feature interactions. 

Recent research progressively explores side information fusion methods. SASRec$_F$ \cite{DIF-SR:conf/sigir/XieZK22} chooses to merge side information before model input but faces risks of information invasion. NOVA \cite{NOVA:journals/corr/abs-2103-03578} prevents information invasion via a non-invasive attention mechanism---using attribute information solely to compute attention weights while maintaining ID embeddings as independent value matrices,  as illustrated in Fig.~\ref{fig3:dif}(a). DIF-SR \cite{DIF-SR:conf/sigir/XieZK22} enables personalized learning of side information through decoupled attention matrices at the attention layer, supporting high-rank mapping and gradient optimization, as shown in Fig.~\ref{fig3:dif}(b). DLFS-Rec \cite{DLFS-Rec:conf/recsys/LiuD0P023} builds side information-guided frequency-domain filters enhanced by uncertainty modeling. MSSR \cite{MSSR:conf/wsdm/LinLPP0LH024} optimizes representations via multi-sequence cross attention as shown in Fig.~\ref{fig3:dif}(c), as well as self-supervised alignment. ASIF \cite{ASIF:conf/www/WangSMHZZZM24} first decouples positional encoding to eliminate noise, and then aligns ID embedding and attribute embedding through contrastive learning, ultimately extracting homogeneous features via orthogonal decomposition. DIFF~\cite{DIFF:conf/sigir/Kim00BL25} introduces a dual-fusion architecture that employs a frequency-domain filter for input denoising and integrates both early and intermediate fusion streams to achieve deep utilization of side information, as illustrated in Fig.~\ref{fig3:dif}(d).

\section{PRELIMINARY}
In this section, we first formulate the problem of sequential recommendation with side information fusion and then introduce the time span partitioning mechanism.

\subsection{Problem Formulation}
Let $\mathcal{U}$ denote the set of users and $\mathcal{I}$ denote the set of items. The universe of item attributes is represented by a set of attribute types $\mathcal{A} = \{A_1, A_2, \dots, A_{|\mathcal{A}|}\}$, such as \textit{category} or \textit{brand}. For each attribute type $A_j \in \mathcal{A}$, there is a corresponding set of discrete attribute values, denoted by $\mathcal{A}_j$. Consequently, each item $i \in \mathcal{I}$ is characterized by a unique identifier \texttt{id} and a tuple of attribute values $(a_{i,1}, a_{i,2}, \dots, a_{i,|\mathcal{A}|})$, where $a_{i,j} \in \mathcal{A}_j$.

For each user $u \in \mathcal{U}$, their historical interactions form a chronologically ordered sequence $S_u = [(i_1, t_1), (i_2, t_2), \dots, (i_{|S_u|}, t_{|S_u|})]$. Each tuple $(i_k, t_k)$ signifies that user $u$ interacted with item $i_k \in \mathcal{I}$ at a specific timestamp $t_k$.

The task of sequential recommendation is to predict the next item that user $u$ will most likely interact with, given their historical sequence $S_u$. Formally, the objective is to find an item $i^*$ that maximizes the conditional probability: $i^* = \arg\max_{i \in \mathcal{I}} P(i | S_u)$.

\subsection{Time Span Partitioning}
We present a time span partitioning mechanism, as illustrated in Figure~\ref{fig1: TSP}. First, we discretize the global timeline into $N_\tau$ disjoint and uniform intervals, which we term \textit{time spans} (e.g., month, quarter). Each span is assigned a unique, learnable time token $\tau \in \{\tau_1, \dots, \tau_{N_\tau}\}$. For each user interaction sequence $S_u$, we then generate a parallel time token sequence $S_u^\tau = (\tau_{k_1}, \dots, \tau_{k_{|S_u|}})$ by mapping each interaction's timestamp $t_k$ to its corresponding time span's token.

This time span partitioning mechanism offers significant benefits. By mapping interactions to a discrete set of learnable tokens, it endows the model with explicit temporal awareness. Furthermore, because these tokens are derived from a uniformly partitioned global timeline, they act as a shared temporal anchor. This aligns the temporal contexts across different user sequences, enabling the model to more easily capture periodic patterns in user behavior.

Distinct from approaches that inject temporal tokens directly into the item sequence~\cite{SimpleDyg:conf/www/WuFL24}, our method of maintaining two separate, parallel sequences for items and time tokens offers two key advantages. First, it preserves the structural integrity of the original item sequence, avoiding potential disruption of its learned patterns. Second, it ensures modularity, allowing the mechanism to function as a plug-and-play component for mainstream recommendation architectures without altering their fundamental input structures.

\section{METHODOLOGY}
\begin{figure*}[t]
  \centering
   \includegraphics[width=0.9\linewidth]{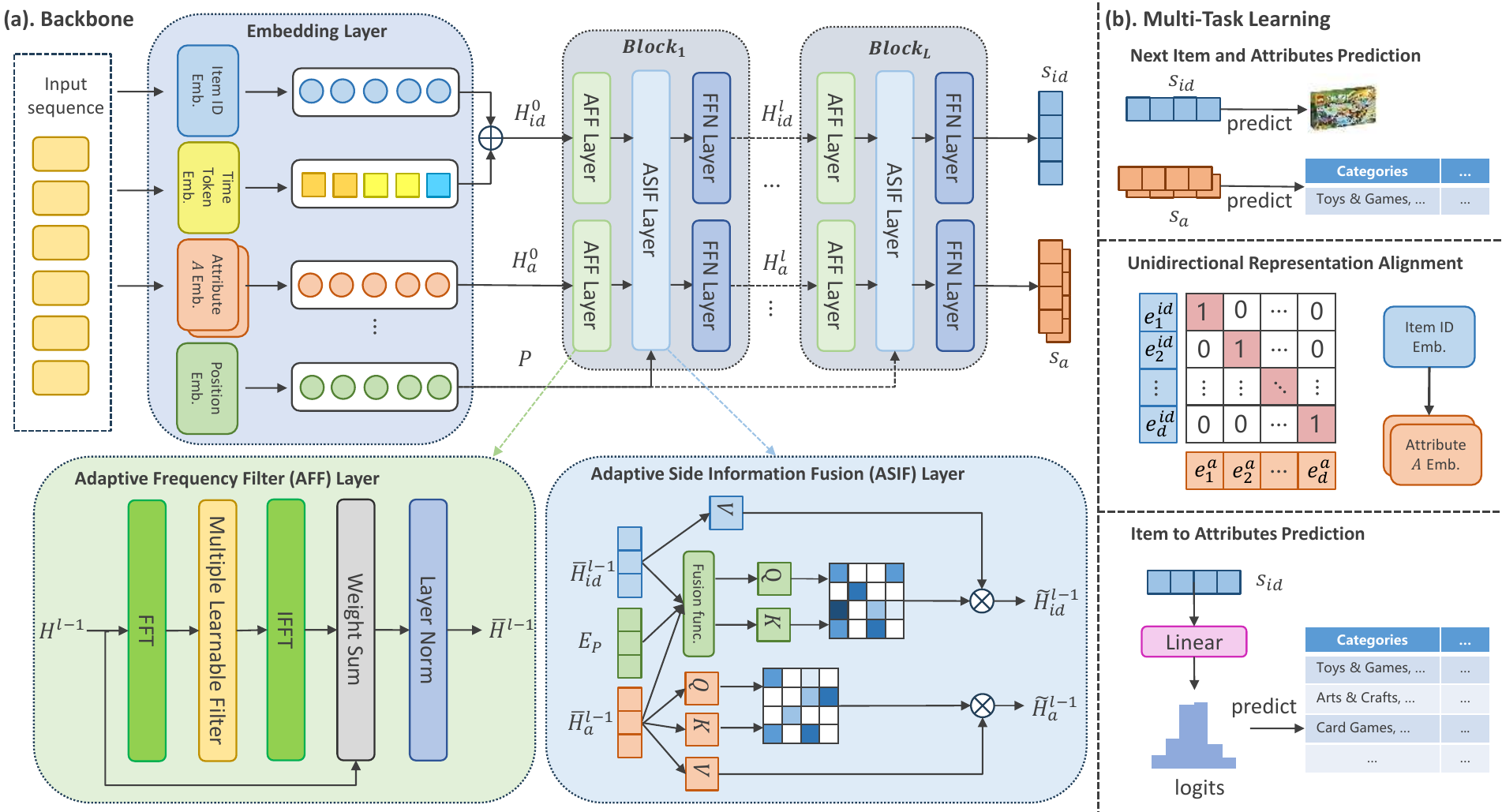}
   \caption{Overall framework of TASIF.}
   \Description{}
   \label{fig4:Overall framework}
\end{figure*}

This section introduces our proposed \textbf{TASIF} framework. As shown in Figure~\ref{fig4:Overall framework}, TASIF first embeds input sequences into ID, time, attribute, and positional embeddings. These are fed into $L$ processing blocks, each comprising an adaptive frequency filter (Section~\ref{Sec: filter}), an adaptive side information fusion layer (Section~\ref{Sec: fusion}), and a feed-forward layer. The final optimized representations are used for model training and prediction (Section \ref{Sec: learning}).

\subsection{Embedding Layer}
The embedding layer is responsible for transforming the discrete item and token sequences into continuous, low-dimensional vector representations. We define the following embedding matrices: item ID embeddings $\mathbf{E}_{ID} \in \mathbb{R}^{|\mathcal{I}| \times d}$, attribute embeddings $\mathbf{E}_{A_j} \in \mathbb{R}^{|\mathcal{A}_j| \times d}$ for each attribute type $A_j \in \mathcal{A}$, time token embeddings $\mathbf{E}_{T} \in \mathbb{R}^{N_\tau \times d}$, and positional embeddings $\mathbf{E}_{P} \in \mathbb{R}^{n \times d}$, where $d$ is the dimensionality of the embeddings and $n$ is the maximum sequence length.

Given a user's interaction sequence $S_u$ and its corresponding time token sequence $S_u^\tau$, we first perform embedding lookups to obtain the initial vector sequences. The primary item representation, $\mathbf{H}_{id}^0$, is constructed by fusing item ID and time token information:
\begin{equation}
    \mathbf{H}_{id}^0 = \text{LayerNorm}(\mathbf{E}_{ID}(S_u) + \mathbf{E}_{T}(S_u^\tau))
    \label{eq:h_init}
\end{equation}
where $\mathbf{E}_{ID}(S_u)$ and $\mathbf{E}_{T}(S_u^\tau)$ represent the sequences of item ID and time token embeddings, respectively. A dropout layer is subsequently applied to $\mathbf{H}_{id}^{0}$ for regularization.


Similarly, for each attribute type $A_j \in \mathcal{A}$, we generate a corresponding attribute embedding sequence: $\mathbf{H}_{a_j}^0 = \mathbf{E}_{A_j}(S_u)$.

\subsection{Adaptive Information Filter and Fusion}
This module consists of multiple Transformer blocks. Each Transformer block includes a Adaptive Frequency Filter (AFF) Layer, an Adaptive Side Information Fusion (ASIF) Layer, and a Feed-Forward Network (FFN) Layer.

\subsubsection{Adaptive Frequency Filter Layer}\label{Sec: filter}
Noisy user behavior sequences negatively impact sequential recommendation, especially in attention-based models \cite{1DBLP:conf/sigir/WuWF0CLX21}. To address this, recent methods incorporate frequency domain analysis to denoise item representations \cite{2DBLP:conf/nips/ZhouMWW0YY022,3DBLP:conf/iclr/XuZ024,FEARec:conf/sigir/DuYZQZ0LS23}. FMLP-Rec \cite{FMLP-Rec:conf/www/ZhouYZW22} and DLFS-Rec \cite{DLFS-Rec:conf/recsys/LiuD0P023} use learnable filters for noise suppression, while BSARec \cite{BSARec:conf/aaai/Shin0WP24} and DIFF~\cite{DIFF:conf/sigir/Kim00BL25} separates signals into low- and high-frequency components, preserving the former and refining the latter. Oracle4Rec \cite{Oracle4Rec:conf/wsdm/XiaLGLZSG25} takes a non-parametric route, directly discarding high-frequency noise.

Considering that treating high-frequency components simply as noise might be oversimplified, we adopt learnable filters and introduce corresponding enhancements to more effectively suppress noise. Specifically, we introduce an adaptive frequency filter layer before each adaptive side information fusion layer, which abandons the original residual connection structure and instead adopts an adaptive weighting approach to merge original features with frequency-domain processed features. At the $l$-th block, we first transform the input hidden states \( H^{l-1} \) from the time domain to the frequency domain using the Fast Fourier Transform (FFT):
\begin{equation}
H_x^{l-1} = \text{FFT}(H^{l-1}),
\end{equation}
where \( H^{l-1} \) collectively represents the hidden states \( H_{id}^{l-1} \) and \( H_{a_j}^{l-1} \), and \( H_x^{l-1} \) is a complex tensor representing the spectrum of \( H^{l-1} \) in the frequency domain. Next, a learnable filter \( W \in \mathbb{R}^{n \times d} \) is applied to modulate the spectrum:
\begin{equation}
\tilde{H}_x^{l-1} = W \odot H_x^{l-1},
\end{equation}
where \( \odot \) denotes element-wise multiplication. Afterward, the modulated spectrum \( \tilde{H}_x^{l-1} \) is transformed back to the time domain using the Inverse Fast Fourier Transform (IFFT), with dropout applied to the resulting features. A learnable weight parameter \( \alpha \) (\( 0 < \alpha < 1 \)) is introduced to dynamically adjust the retention ratio of the frequency-domain filtered features. Finally, the denoised embedding is obtained through layer normalization:
\begin{equation}
\bar{H}^{l-1} = \text{LayerNorm}(\alpha \cdot \text{Dropout}( \text{IFFT}(\tilde{H}_x^{l-1})) + (1 - \alpha) \cdot H^{l-1}).
\label{eq:5}
\end{equation}

This design allows the model to adaptively adjust the intensity of frequency-domain filtering based on different scenarios, suppressing noise while preserving useful high-frequency features.

\subsubsection{Adaptive Side Information Fusion Layer}\label{Sec: fusion}
The Adaptive Side Information Fusion (ASIF) layer unifies deep interaction and computational efficiency through an asymmetric ``guide-not-mix'' architecture. In this design, Query (Q) and Key (K) vectors are generated from a fusion of item ID, attributes and position information to guide the model's attention, whereas the Value (V) vector is derived exclusively from the item IDs to 
preserve the collaborative signal during updates. Within two parallel pathways, attribute representations are also refined via independent self-attention. Consequently, the ASIF layer fully leverages side information to optimize ID sequence modeling while avoiding feature dilution and maintaining computational efficiency.

Specifically, at the $l$-th block, the ASIF layer takes the denoised item hidden state $\bar{H}_{id}^{l-1}$ and attribute hidden state $\bar{H}_{a_j}^{l-1}$ from the preceding AFF layer as input.
First, we fuse the item ID representation, attribute representations, and position embeddings to obtain a unified representation, $H_f^{l-1}$, which is used to generate the attention scores:
\begin{equation}\label{eq:fusion}
H_f^{l-1} = \text{Fusion}(\bar{H}_{id}^{l-1}, \bar{H}_{a_j}^{l-1}, E_P),
\end{equation}
where $\text{Fusion(·)}$ represents the fusion operation, such as concatenation followed by a linear layer.

Next, we use this fused representation $H_f^{l-1}$ to compute the Query and Key matrices, while the Value matrix is generated directly from the original item ID representation $\bar{H}_{id}^{l-1}$. This separation achieves what we term the ``guide-not-mix'' design, as follows.
\begin{equation}
Q_{id} = H_f^{l-1} W_Q^{id}, \quad K_{id} = H_f^{l-1} W_K^{id}, \quad V_{id} = \bar{H}_{id}^{l-1} W_V^{id},
\end{equation}
where $W_Q^{id}, W_K^{id}, W_V^{id} \in \mathbb{R}^{d \times d}$ are learnable projection matrices for the ID pathway.

Finally, the updated item ID hidden state $\tilde{H}_{id}^{l-1}$ is obtained via the standard attention calculation:
\begin{equation}
\tilde{H}_{id}^{l-1} = \text{Softmax}\left( \frac{Q_{id} K_{id}^{\top}}{\sqrt{d}} \right) V_{id}.
\end{equation}

Meanwhile, each attribute representation $\bar{H}_{a_j}^{l-1}$ is updated separately through a parallel, standard self-attention mechanism to refine its own intra-sequence dependencies:
\begin{equation}
Q_{a_j} = \bar{H}_{a_j}^{l-1} W_Q^{a_j}, \quad K_{a_j} = \bar{H}_{a_j}^{l-1} W_K^{a_j}, \quad V_{a_j} = \bar{H}_{a_j}^{l-1} W_V^{a_j},
\end{equation}
\begin{equation}
\tilde{H}_{a_j}^{l-1} = \text{Softmax}\left( \frac{Q_{a_j} K_{a_j}^{\top}}{\sqrt{d}} \right) V_{a_j},
\end{equation}
where $W_Q^{a_j}, W_K^{a_j}, W_V^{a_j} \in \mathbb{R}^{d \times d}$ are learnable projection matrices for the attribute pathway.

The updated representations $\tilde{H}_{id}^{l-1}$ and $\tilde{H}_{a_j}^{l-1}$ are then passed to the subsequent Feed-Forward Network (FFN) layer.

\subsubsection{Feed-Forward Network Layer}
After obtaining \( \tilde{H}_{id}^{l-1} \) and \( \tilde{H}_{a_{j}}^{l-1} \), we apply the Feed-Forward Network Layers to obtain the output of the $l$-th block,  \( H_{id}^{l} \) and \(\{H_{a_j}^{l}\}_{j=1}^{|\mathcal{A}|}\):
\begin{equation}
H_{id}^{l} = \text{FFN}(\tilde{H}_{id}^{l-1}), \quad
H_{a_j}^{l} = \text{FFN}_{a_j}(\tilde{H}_{a_j}^{l-1}).
\end{equation}
Here, FFN refers to the fully connected Feed-Forward Network.

\subsection{Multi-Task Learing and Prediction}\label{Sec: learning}
After the $L$-th (last) block, we extract the final hidden states of the ID stream, \( H_{id}^L \), and each attribute stream, \(\{H_{a_j}^L\}_{j=1}^{|\mathcal{A}|}\). These states, denoted by \( s_{id} \in \mathbb{R}^{d}\) and \( \{s_{a_j} \in \mathbb{R}^{d}\}_{j=1}^{|\mathcal{A}|}\), serve as the user's final interest representations for downstream tasks.

\subsubsection{Primary Recommendation Task}
The primary task is to predict the next item and its attributes. We compute the predictive distributions via inner product, 
followed by a Softmax layer:
\begin{equation}
\hat{y}_{id} = \text{Softmax}(s_{id} E_{ID}^{\top}), \quad \hat{y}_{a_j} = \text{Softmax}(s_{a_j}, E_{A_j}^{\top})
\end{equation}
where \( E_{ID} \) and \( E_{A_j} \) are the embedding matrices for the item IDs and the \(j\)-th attribute type, respectively. We then train the model using the cross-entropy loss against the ground-truth labels. The recommendation losses are formulated as:
\begin{equation}
\mathcal{L}_{\text{rec}}^{ID} = -\sum_{i=1}^{B} y_{id}^{(i)} \log(\hat{y}_{id}^{(i)}), \quad
\mathcal{L}_{\text{rec}}^{A_j} = -\sum_{i=1}^{B} y_{a_j}^{(i)} \log(\hat{y}_{a_j}^{(i)}),
\end{equation}
where \(B\) is the batch size, \(y_{id}^{(i)}\) and \(y_{a_j}^{(i)}\) are the one-hot encoded ground-truth labels for the \(i\)-th sample in the batch.

\subsubsection{Unidirectional Representation Alignment}
Inspired by Wang \emph{et al.}~\cite{ASIF:conf/www/WangSMHZZZM24}, we align ID and attribute embeddings. To account for the varying importance of attributes, we compute a separate loss for each type and combine them via a weighted sum. Furthermore, we recognize that since an item's attributes can be non-unique, a bidirectional alignment would risk introducing conflicting supervisory signals.
Therefore, we adopt a unidirectional strategy, driving the alignment only from the unique ID embedding $e^{id}$ towards its corresponding attribute embedding $e^{a_j}$.
This alignment process is driven by the InfoNCE loss~\cite{InfoNCE:journals/corr/abs-1807-03748}, which aims to pull positive pairs closer while pushing negative pairs apart:
\begin{equation}
    \mathcal{L}_{\text{align}}^{A_j} = -\log \frac{\exp(\text{sim}(e_{i}^{id}, e_{i}^{a_j})/\tau)}{\sum_{k=1}^{B} \exp(\text{sim}(e_{i}^{id}, e_{k}^{a_j})/\tau)},
    \label{eq:infonce}
\end{equation}
where \(\text{sim}(\cdot, \cdot)\) is cosine similarity, \(B\) is the batch size and \(\tau\) is a temperature hyperparameter.

\subsubsection{Item-to-Attribute Prediction}
Following Xie \emph{et al.}~\cite{DIF-SR:conf/sigir/XieZK22}, to further enrich user's ID-based representation \(s_{id}\), we add an auxiliary task to predict the target item's attributes from \(s_{id}\). The predicted probabilities for the values of the \(j\)-th attribute type are given by:
\begin{equation}
\hat{y}_{i2a}^{A_j} = \text{Sigmoid}(W_{A_j} s_{id}^{\top} + b_{A_j}),
\end{equation}
where \( W_{A_j} \in \mathbb{R}^{|\mathcal{A}_j| \times d} \) and \( b_{A_j} \in \mathbb{R}^{|\mathcal{A}_j|} \) are learnable parameters. We then compute the binary cross-entropy (BCE) loss for this multi-label classification task:
\begin{equation}
\mathcal{L}_\text{i2a}^{A_j} = -\sum_{k=1}^{|\mathcal{A}_j|} \left[ y_{i2a, k}^{A_j} \log(\hat{y}_{i2a, k}^{A_j}) + (1 - y_{i2a, k}^{A_j}) \log(1 - \hat{y}_{i2a, k}^{A_j}) \right]
\end{equation}
where \(y_{i2a, k}^{A_j}\) is the ground-truth label for the \(k\)-th value of the \(j\)-th attribute type.

\subsubsection{Joint Optimization}
The final objective is a weighted sum of all losses. The overall loss \(\mathcal{L}\) is defined as:
\begin{equation}
\mathcal{L} = \mathcal{L}_{\text{rec}}^{ID} + \sum_{j=1}^{|\mathcal{A}|} W_j \left( \lambda_1 \mathcal{L}_{\text{rec}}^{A_j} + \lambda_2 \mathcal{L}_{\text{align}}^{A_j} + \lambda_3 \mathcal{L}_{\text{i2a}}^{A_j} \right)
\end{equation}
Here, \( \lambda_k \) are fixed hyperparameters balancing the tasks, while \(W_j\) are learnable weights that allow the model to dynamically assess the importance of each attribute type.

\subsubsection{Prediction}
In the prediction phase, the model generates recommendations by comprehensively considering both item IDs and their associated side information (attributes). Specifically, the ID-based interaction score is computed via the dot product of the final user interest representations for IDs, $s_{id}$, and the item ID embedding matrix, $E_{ID}$. In parallel, interaction scores for each attribute type $j$ are derived from their respective interest representations, $s_{a_j}$, and embedding matrices, $E_{A_j}$.
Crucially, these attribute scores are first scaled by their importance weights $W_j$, which were learned during the training phase. The final composite score is then formed as a weighted combination of the ID and aggregated attribute scores, controlled by the hyperparameter $\beta$:
\begin{equation}
\text{score} = (1 - \beta) s_{id} E_{ID}^{\top} + \beta \sum_{j=1}^{|\mathcal{A}|} W_j s_{a_j} E_{A_j}^{\top}.
\label{eq:predict_score}
\end{equation}
Here, the hyperparameter $\beta \in (0, 1)$ balances the trade-off between the primary ID-based predictions and the attribute-informed predictions. Finally, items are ranked based on \( \text{score} \), and the item with the highest score is recommended to the user.

\subsection{Efficiency Analysis}
We analyze the computational complexity by comparing our model, TASIF, with MSSR, focusing on the fusion layers where the two models differ most. Let $d$ be the embedding dimension, $n$ the sequence length, and $|\mathcal{A}|$ the number of attributes. The computational complexity of a standard Transformer self-attention layer serves as our analytical baseline, with complexity $O(n^2 \cdot d + n \cdot d^2)$, where $O(n^2 \cdot d)$ accounts for attention computations (Query-Key multiplication and Attention-Value aggregation) and $O(n \cdot d^2)$ represents the linear projection overhead for Q, K, V generation and output transformation. Our model, TASIF, computes one main attention stream and $|\mathcal{A}|$ independent attribute streams in parallel. This results in a total complexity that scales linearly with the number of attributes: $\mathcal{O}\left((|\mathcal{A}+1|)(n d^2 + n^2 d)\right)$. In contrast, MSSR utilizes a score-level cross-fusion strategy, which explicitly computes cross-attention for all pairs among its $|\mathcal{A}|+2$ attributes (including  position). This requires computing $(|\mathcal{A}|+2)^2$ attention matrices, leading to a total complexity of $\mathcal{O}\left((|\mathcal{A}|+2)^2 n^2 d + (|\mathcal{A}| + 2) n d^2 \right)$. Consequently, MSSR's complexity exhibits quadratic growth with the number of attributes $|\mathcal{A}|$, while TASIF maintains linear scaling. This demonstrates the superior computational efficiency of our approach, which is empirically validated in Section~\ref{Sec: efficiency}.

\section{EXPERIMENT}
In this section, we conduct extensive experiments to verify the effectiveness of our TASIF.

\begin{table}[t]
    \centering
    \caption{Statistics of the datasets after preprocessing.}
    \label{tab1:dataset_statistics}
    \begin{tabular}{lcccc}
        \toprule
        Dataset & Yelp & Beauty & Sports & Toys \\
        \midrule
        \# Users & 30,499 & 22,363 & 35,598 & 19,412 \\
        \# Items & 20,068 & 12,101 & 18,357 & 11,924 \\
        \# Avg. Actions / User & 10.4 & 8.9 & 8.3 & 8.6 \\
        \# Avg. Actions / Item & 15.8 & 16.4 & 16.1 & 14.1 \\
        \# Actions & 317,182 & 198,502 & 296,337 & 167,597 \\
        Sparsity & 99.95\% & 99.93\% & 99.95\% & 99.93\% \\
        \bottomrule
    \end{tabular}
\end{table}
\begin{table*}[t]
\centering
\setlength{\tabcolsep}{1mm}
\caption{Performance comparison of TASIF and the baselines. \textbf{Bold} and \underline{underline} denote the best and second-best results, respectively. \textit{Improv.} denotes the improvement of TASIF over the best baseline.  * indicates that the improvement over the best baseline is statistically significant according to paired t-tests ($p$ < 0.01). $^\dagger$ marks results obtained from literature due to unavailable source code. R@K and N@K represent Recall@K and NDCG@K, respectively.}
\resizebox{\textwidth}{!}{
    \begin{tabular}{c|cccc|cccc|cccc|cccc}
    \hline
    \multicolumn{1}{l|}{\multirow{2}{*}{Method}} & \multicolumn{4}{c|}{Yelp} & \multicolumn{4}{c|}{Beauty} & \multicolumn{4}{c|}{Sports} & \multicolumn{4}{c}{Toys} \\ \cline{2-17} 
    \multicolumn{1}{c|}{} & \multicolumn{1}{c}{R@10} & \multicolumn{1}{c}{R@20} & \multicolumn{1}{c}{N@10} & \multicolumn{1}{c|}{N@20} & \multicolumn{1}{c}{R@10} & \multicolumn{1}{c}{R@20} & \multicolumn{1}{c}{N@10} & \multicolumn{1}{c|}{N@20} & \multicolumn{1}{c}{R@10} & \multicolumn{1}{c}{R@20} & \multicolumn{1}{c}{N@10} & \multicolumn{1}{c|}{N@20} & \multicolumn{1}{c}{R@10} & \multicolumn{1}{c}{R@20} & \multicolumn{1}{c}{N@10} & \multicolumn{1}{c}{N@20}\\ \hline
    GRU4Rec & 0.0382 & 0.0632 & 0.0194 & 0.0256 & 0.0614 & 0.0908 & 0.0333 & 0.0407 & 0.0314 & 0.0491 & 0.0164 & 0.0209 & 0.0422 & 0.0642 & 0.0226 & 0.0281 \\
    SASRec & 0.0597 & 0.0864 & 0.0361 & 0.0428 & 0.0842 & 0.1192 & 0.0416  & 0.0504 & 0.0480 & 0.0703 & 0.0225 & 0.0282 & 0.0868 & 0.1193 & 0.0435 & 0.0517 \\
    TiSASRec & 0.0621 & 0.0887 & 0.0378 & 0.0444 & 0.0859 & 0.1221 & 0.0424 & 0.0519 & 0.0497 & 0.0712 & 0.0238 & 0.0292 & 0.0902 & 0.1212 & 0.0451 & 0.0528 \\
    FMLP-Rec & 0.0671 & 0.0975 & 0.0396 & 0.0471 & 0.0864 & 0.1198 & 0.0432 & 0.0517 & 0.0514 & 0.0735 & 0.0244 & 0.0301 & 0.0902 & 0.1249 & 0.0445 & 0.0534 \\
    BSARec & 0.0701 & 0.1004 & 0.0423 & 0.0499 & 0.0869 & 0.1226 & 0.0435 & 0.0522 & 0.0510 & 0.0741 & 0.0235 & 0.0292 & 0.0942 & 0.1277 & 0.0460 & 0.0544 \\ \hline
    GRU4Rec$_F$ & 0.0320 & 0.0519 & 0.0161 & 0.0212 & 0.0598 & 0.0890 & 0.0322 & 0.0396 & 0.0333 & 0.0535 & 0.0167 & 0.0218 & 0.0506 & 0.0768 & 0.0270 & 0.0336 \\ 
    SASRec$_F$ & 0.0420 & 0.0672 & 0.0223 & 0.0287 & 0.0696 & 0.1020 & 0.0401 & 0.0483 & 0.0424 & 0.0640 & 0.0233 & 0.028& 0.0707 & 0.1033 & 0.0399 & 0.0481 \\
    S$^3$Rec & 0.0671 & 0.0985 & 0.0401 & 0.0480 & 0.0849 & 0.1220 & 0.0422 & 0.0515 & 0.0526 & 0.0778 & 0.0246 & 0.0309 & 0.0940 & 0.1304 & 0.0476 & 0.0568 \\
    FDSA & 0.0534 & 0.0816 & 0.0289 & 0.0360 & 0.0763 & 0.1093 & 0.0461 & 0.0545 & 0.0476 & 0.0675 & 0.0285 & 0.0335 & 0.0881 & 0.1164 & 0.0497 & 0.0563 \\
    NOVA & 0.0669 & 0.0975 & 0.0407 & 0.0484 & 0.0879 & 0.1240 & 0.0442 & 0.0533 & 0.0518 & 0.0757 & 0.0243 & 0.0303 & 0.0965 & 0.1332 & 0.0476 & 0.0568 \\
    DIF-SR & 0.0679 & 0.0999 & 0.0412 & 0.0492 & 0.0900 & 0.1265 & 0.0445 & 0.0536 & 0.0553 & 0.0802 & 0.0258 & 0.0321 & 0.1010 & 0.1377 & 0.0506 & 0.0598 \\
    DLFS-Rec & 0.0652 & 0.0943 & 0.0398 & 0.0471 & 0.0841 & 0.1203 & 0.0424 & 0.0515 & 0.0489 & 0.0729 & 0.0230 & 0.0291 & 0.0907 & 0.1225 & 0.0475 & 0.0555 \\
    MSSR & 0.0713 & 0.1043 & 0.0427 & 0.0511 & 0.0901 & 0.1293 & 0.0446 & 0.0545 & 0.0552 & 0.0816 & 0.0262 & 0.0329 & \underline{0.1017} & \underline{0.1405} & \underline{0.0507} & \underline{0.0605} \\
    ASIF$^\dagger$ & 0.0724 & 0.1052 & 0.0427 & 0.0510 & \underline{0.0920} & \underline{0.1322} & \underline{0.0463} & 0.0564 & \underline{0.0568} & \underline{0.0827} & 0.0268 & 0.0333 & 0.1007 & 0.1393 & 0.0496 & 0.0593 \\
    DIFF & \underline{0.0811} & \underline{0.1175} & \underline{0.0467} & \underline{0.0558} & 0.0856 & 0.1221 & \textbf{0.0491} & \textbf{0.0582} & 0.0527 & 0.0783 & \textbf{0.0297} & \underline{0.0359} & 0.0928 & 0.1322 & \underline{0.0507} & 0.0603 \\ \hline
    TASIF & \textbf{0.0862*} & \textbf{0.1251*} & \textbf{0.0488*} & \textbf{0.0586*} & \textbf{0.0949*} & \textbf{0.1373*} & 0.0462 & \underline{0.0568} & \textbf{0.0611*}  & \textbf{0.0912*} & \underline{0.0288} & \textbf{0.0363} & \textbf{0.1084*} & \textbf{0.1511*} & \textbf{0.0529*} & \textbf{0.0636*} \\
    \textit{Improv.} & 6.29\% & 6.47\% & 4.50\% & 5.02\% & 3.15 \%& 3.86\% & - & - & 7.57\% & 10.28\% & - & 1.11\% & 6.59\% & 7.54\% & 4.34\% & 5.12\% \\ \hline
    \end{tabular}
}
\label{tab2:overall}
\end{table*}

\subsection{Experimental Settings}
\subsubsection{Dataset}
We conduct experiments on four widely used real-world datasets, including Yelp\footnote{\url{https://www.yelp.com/dataset}} and three datasets constructed from Amazon review data\footnote{\url{http://jmccauley.ucsd.edu/data/amazon/}} \cite{Amazon:conf/sigir/McAuleyTSH15}: Beauty, Sports, and Toys. Yelp is a publicly available dataset for business recommendations, while the Beauty, Sports, and Toys datasets originate from the Amazon review dataset, containing detailed user evaluations of products. For consistency with earlier research \cite{DIF-SR:conf/sigir/XieZK22, MSSR:conf/wsdm/LinLPP0LH024, ASIF:conf/www/WangSMHZZZM24}, we use category and position information as attribute information in all datasets. During data preprocessing, we follow the strategy of Xie \emph{et al.}~\cite{DIF-SR:conf/sigir/XieZK22}, removing users and items that appear fewer than five times and applying the leave-one-out strategy to split the training, validation, and test sets. Additionally, for the Yelp dataset, we retain only transaction records after January 1, 2019 \cite{DIF-SR:conf/sigir/XieZK22}. Table~\ref{tab1:dataset_statistics} summarizes the statistics of the processed datasets.

\subsubsection{Evaluation Metrics}
We utilize Recall@K and NDCG@K (\( K \in \{10, 20\} \)) as evaluation metrics, where Recall@K measures the proportion of relevant items appearing in the recommendation list, while NDCG@K considers the ranking quality of those items. To ensure fairness, we adopt a full ranking evaluation setup~\cite{Metrics:conf/kdd/KricheneR20} rather than relying on sampled items, providing a more accurate assessment of model performance. All results are reported as the average over five runs with different random seeds.

\subsubsection{Baseline Models}
Baselines are categorized into models without and with side information.

Models without side information:
(1) \textbf{GRU4Rec} \cite{GRU4Rec:journals/corr/HidasiKBT15} models sequential user behavior patterns using GRU networks;
(2) \textbf{SASRec} \cite{SASRec:conf/icdm/KangM18} learns user representations from item sequences using self-attention;
(3) \textbf{TiSASRec} \cite{TiSASRec:conf/wsdm/LiWM20} captures temporal dynamics through time-interval aware self-attention;
(4) \textbf{FMLP-Rec} \cite{FMLP-Rec:conf/www/ZhouYZW22} combines frequency domain filtering and MLPs to enhance sequential recommendation;
(5) \textbf{BSARec} \cite{BSARec:conf/aaai/Shin0WP24} separates and fuses frequency features via Fourier transforms to mitigate self-attention over-smoothing.

Models with side information:
(1) \textbf{GRU4Rec$_F$} enhances GRU4Rec by fusing item and side information;
(2) \textbf{SASRec$_F$} extends SASRec by incorporating item and side information;
(3) \textbf{FDSA} \cite{FDSA:conf/ijcai/ZhangZLSXWLZ19} processes item and feature sequences through separate self-attention modules with late fusion;
(4) \textbf{S$^3$Rec} \cite{S3Rec:conf/cikm/ZhouWZZWZWW20} designs multi-task contrastive pre-training based on mutual information maximization;
(5) \textbf{NOVA} \cite{NOVA:journals/corr/abs-2103-03578} proposes non-invasive attention mechanism for user representation learning;
(6) \textbf{DIF-SR} \cite{DIF-SR:conf/sigir/XieZK22} decouples attention computation using fusion attention for user representation learning;
(7) \textbf{DLFS-Rec} \cite{DLFS-Rec:conf/recsys/LiuD0P023} leverages side information-guided learnable filters for frequency-domain denoising;
(8) \textbf{MSSR} \cite{MSSR:conf/wsdm/LinLPP0LH024} models correlations between item sequence and feature sequences;
(9) \textbf{ASIF} \cite{ASIF:conf/www/WangSMHZZZM24} optimizes feature fusion through representation alignment and homogeneous information extraction;
(10) \textbf{DIFF}~\cite{DIFF:conf/sigir/Kim00BL25} denoises inputs via a frequency-domain filter and uses parallel early and intermediate fusion streams to deeply integrate side information.

\subsubsection{Implementation Details}
To ensure a fair and reproducible comparison, all baseline models and our proposed method are implemented within the RecBole framework~\cite{RecBole:conf/cikm/ZhaoMHLCPLLWTMF21}. We uniformly train all models using the Adam optimizer~\cite{Adam:journals/corr/KingmaB14} with a learning rate of 0.0001. For the baselines, we tuned their hyperparameters based on the search spaces recommended in their original papers. For our proposed method, the hidden state dimension is set to 256. The batch size is configured to 2048 for the three Amazon datasets and 1024 for the Yelp dataset. The number of transformer layers and attention heads follow the settings in Xie \emph{et al.}~\cite{DIF-SR:conf/sigir/XieZK22}. We perform a grid search for the time span $\tau$ over $\{7, 30, 90, 180, 365\}$ days and for the attribute score weight $\beta$ in Equation (\ref{eq:predict_score}) over $\{0.1, 0.2, 0.3, 0.4, 0.5\}$. The loss weights are fixed at $\lambda_1=1.0$ and $\lambda_2=0.1$, while $\lambda_3$ is set to 10 for Yelp and Beauty datasets and 5 for Sports and Toys datasets. The $\text{Fusion(·)}$ function in Equation (\ref{eq:fusion}) is set to a gating mechanism for the Yelp and Sports datasets, and to element-wise summation for the Beauty and Toys datasets. All experiments were conducted on a single NVIDIA GeForce RTX 4090 GPU with 24GB of memory.

\subsection{Overall Performance Comparison}
Table ~\ref{tab2:overall} presents performance comparisons across all datasets, revealing several key findings: (1) Among basic sequential recommendation models, TiSASRec outperforms SASRec by incorporating temporal interval information, highlighting the critical role of temporal modeling in sequential recommendation. In addition, BSARec achieves better performance than SASRec by introducing a filtering mechanism, demonstrating that such filters are effective in mitigating noise in user behavior sequences; (2) Regarding the utilization of side information, early simple fusion methods (such as GRU4Rec$_F$ and SASRec$_F$) performed worse than their baseline versions due to the information interference problem. FDSA partially addressed this issue by adopting a feature separation modeling strategy, achieving performance improvements; (3) Recent information fusion approaches, including NOVA, DIF-SR, MSSR and so on, have further enhanced model performance through more sophisticated fusion mechanisms, demonstrating that proper utilization of side information can significantly improve recommendation performance; (4) Notably, our proposed TASIF method achieves state-of-the-art performance across various datasets, significantly outperforming existing methods in most cases. These results validate the effectiveness and superiority of our approach in both side information fusion and temporal modeling.

To further evaluate TASIF’s capability in multi-side information fusion, following Lin \emph{et al.}~\cite{MSSR:conf/wsdm/LinLPP0LH024}, we incorporate additional attributes—“city” for Yelp and “brand” for  Beauty, Sports and Toys. As shown in Table~\ref{tab3:side_info}, TASIF consistently outperforms two representative baselines across all settings on most metrics. Compared with Table~\ref{tab2:overall}, adding side information generally enhances model performance, with TASIF achieving the most notable gains on the Beauty and Toys dataset. These results highlight TASIF’s effectiveness in modeling multiple types of side information.
\begin{table}[!t]
    \centering
    \caption{Performance comparison of our TASIF and two representative baselines while using more side information.}
    \label{tab3:side_info}
    \resizebox{0.48\textwidth}{!}{
        \begin{tabular}{c|c|c|c|c|c}
            \toprule
            Datasets & Model & Recall@10 & Recall@20 & NDCG@10 & NDCG@20 \\        
            \midrule
            \multirow{3}{*}{Yelp} 
            & MSSR & 0.0716 & 0.1069 & 0.0431 & 0.0518 \\
            & DIFF & \underline{0.0816} & \underline{0.1182} & \underline{0.0472} & \underline{0.0563} \\
            & TASIF & \textbf{0.0871} & \textbf{0.1253} & \textbf{0.0491} & \textbf{0.0587} \\
            \midrule
            \multirow{3}{*}{Beauty} 
            & MSSR & 0.0916 & 0.1322 & 0.0463 & 0.0559 \\
            & DIFF & \underline{0.0928} & \underline{0.1346} & \textbf{0.0527} & \textbf{0.0632} \\
            & TASIF & \textbf{0.1024} & \textbf{0.1504} & \underline{0.0506} & \underline{0.0625} \\
            \midrule
            \multirow{3}{*}{Sports} 
            & MSSR & 0.0564 & 0.0819 & 0.0261 & 0.0327 \\
            & DIFF & \underline{0.0572} & \underline{0.0838} & \textbf{0.0312} & \textbf{0.0378} \\
            & TASIF & \textbf{0.0614} & \textbf{0.0937} & \underline{0.0291} & \underline{0.0372} \\
            \midrule
            \multirow{3}{*}{Toys} 
            & MSSR & \underline{0.1035} & 0.1419 & 0.0521 & 0.0614 \\
            & DIFF & 0.1015 & \underline{0.1437} & \underline{0.0542} & \underline{0.0648} \\
            & TASIF & \textbf{0.1121} & \textbf{0.1595} & \textbf{0.0551} & \textbf{0.0670} \\
            \bottomrule
        \end{tabular}
    }
\end{table}

\subsection{Ablation Study}
To precisely evaluate the contribution of each component within TASIF, we conducted systematic ablation studies as presented in Table~\ref{tab4:ablation}. The findings clearly indicate that every module plays an important role. \textit{w/o TSP}, removing the Time Span Partitioning (TSP) mechanism leads to a significant performance degradation, highlighting its crucial role in capturing temporal dynamics. \textit{w/o AFF}, eliminating the Adaptive Frequency Filter (AFF) causes a notable decline, demonstrating the importance of denoising for enhancing feature quality and user representations. \textit{w/o ASIF}, replacing our Adaptive Side Information Fusion (ASIF) layer with the fusion method from a baseline model~\cite{MSSR:conf/wsdm/LinLPP0LH024} results in a substantial performance drop, validating the superiority of our fusion architecture. Finally, for the auxiliary task ablations, \textit{w/o NAP}, \textit{w/o URA}, and \textit{w/o I2A}, removing any of the Next Attribute Prediction (NAP), Unidirectional Representation Alignment (URA), or Item-to-Attribute Prediction (I2A) objectives consistently degrades performance across all datasets. This collectively underscores that robustly learning attribute representations is fundamental to their effective fusion and utilization. In summary, these results demonstrate that each component of TASIF is essential to its overall performance, and their synergy enables state-of-the-art recommendation accuracy.

\begin{table}[!t]
\caption{Ablation study of our TASIF (Recall@20).}
\label{tab4:ablation}
\begin{tabular}{l|cccc}
\hline
Variant & Yelp & Beauty & Sports & Toys \\
\hline
w/o TSP & 0.1212 & 0.1365 & 0.0862 & 0.1442 \\
w/o AFF & 0.1058 & 0.1344 & 0.0869 & 0.1472 \\
w/o ASIF & 0.1163 & 0.1356 & 0.0844 & 0.1447 \\
w/o NAP & 0.1224 & 0.1371 & 0.0908 & 0.1465 \\
w/o URA & 0.1213 & 0.1372 & 0.0887 & 0.1475 \\
w/o I2A & 0.1206 & 0.1369 & 0.0849 & 0.1447 \\
\hline
TASIF & \textbf{0.1251} & \textbf{0.1373} & \textbf{0.0912} & \textbf{0.1511} \\
\hline
\end{tabular}
\end{table}


\subsection{Filter Type Study}
\begin{figure}[t]
  \centering
   \includegraphics[width=\linewidth]{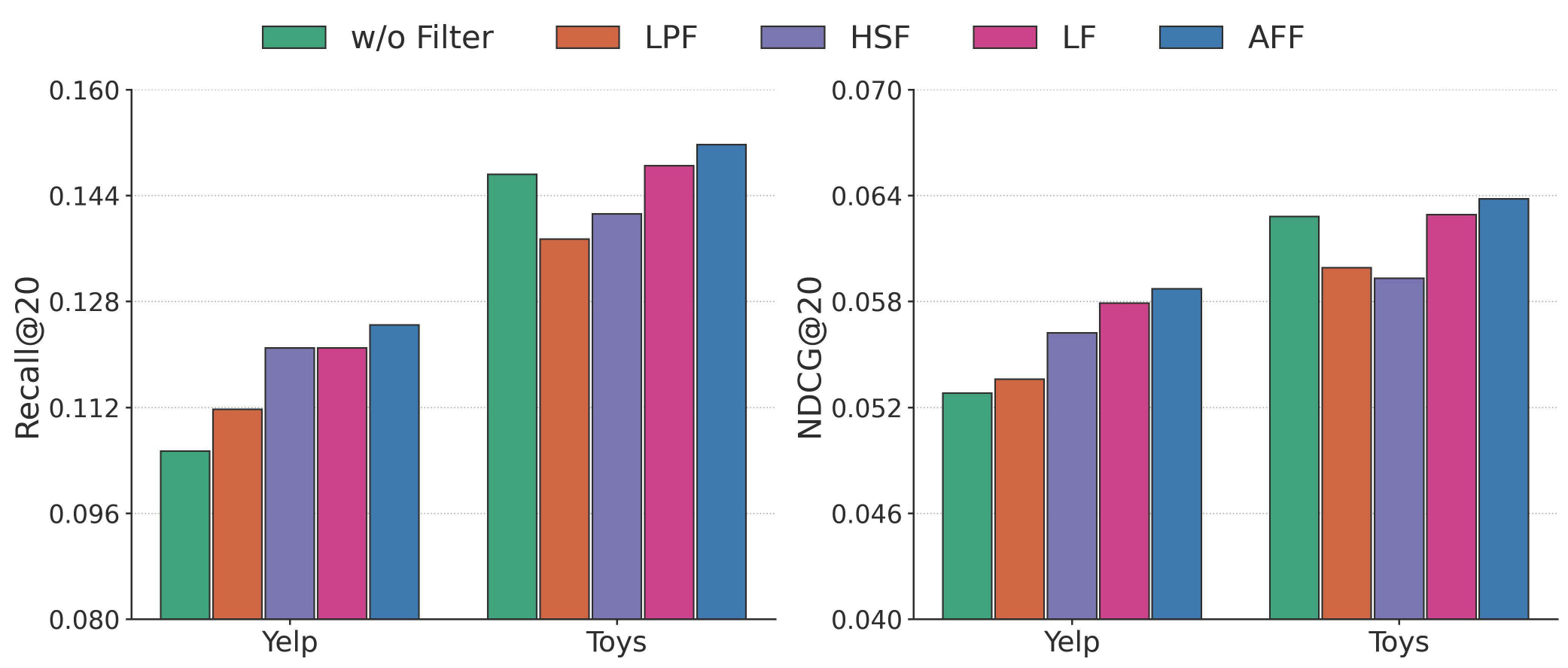}
   \caption{Comparison of different filter types.}
   \Description{}
   \label{fig:5}
\end{figure}

To validate the effectiveness of our proposed Adaptive Frequency Filter (AFF), we conducted additional analysis comparing it with several alternative filter types: a model without any filter (w/o Filter), the Low Pass Filter (LPF)~\cite{Oracle4Rec:conf/wsdm/XiaLGLZSG25}, the High-frequency Suppression Filter (HSF)~\cite{BSARec:conf/aaai/Shin0WP24, DIFF:conf/sigir/Kim00BL25}, and the Learnable Filter (LF)~\cite{FMLP-Rec:conf/www/ZhouYZW22, DLFS-Rec:conf/recsys/LiuD0P023}. The results in Fig.~\ref{fig:5} show that the fixed filters, LBF and HFSF, are sensitive to data scale, performing poorly on the smaller Toys dataset but outperforming the model without a filter on the larger Yelp dataset. In contrast, the learnable filters (LF and AFF) demonstrate stronger generalization and obtain superior performance on both datasets. Notably, our proposed AFF achieves the best results in all settings and shows a consistent performance improvement over LF, which effectively validates the effectiveness of our design.

\subsection{Hyperparameter Study}
We further examine the impact of different hyperparameter settings on model performance.

\begin{figure}[t]
  \centering
   \includegraphics[width=\linewidth]{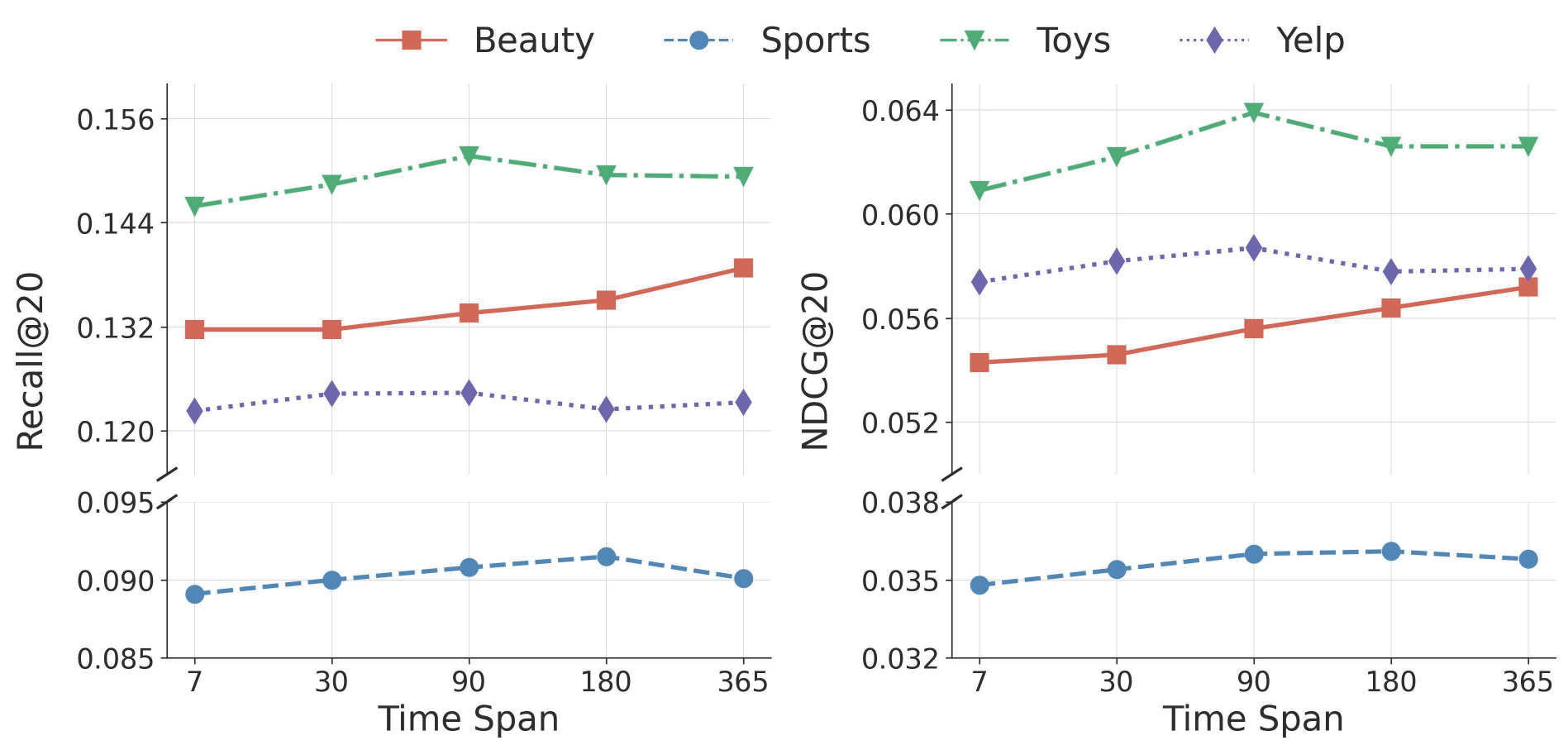}
   \caption{Effects of time span.}
   \Description{}
   \label{fig:6}
\end{figure}

\vspace{1mm}
\noindent\textbf{Time Span}. We first investigated the impact of different time spans (7, 30, 90, 180, and 365 days) on model performance. As shown in Fig.~\ref{fig:6}, the model achieves optimal results with a 90-day span on the Yelp and Toys datasets. For the Sports dataset, the optimal span is 180 days, whereas a 365-day span is best for the Beauty dataset. This divergence can be attributed to the inherent characteristics of the datasets: Yelp (dining services), Toys, and Sports (sporting goods) exhibit strong seasonal patterns, while Beauty (cosmetic products) typically has a longer, near-annual refresh cycle.

\begin{figure}[t]
  \centering
   \includegraphics[width=\linewidth]{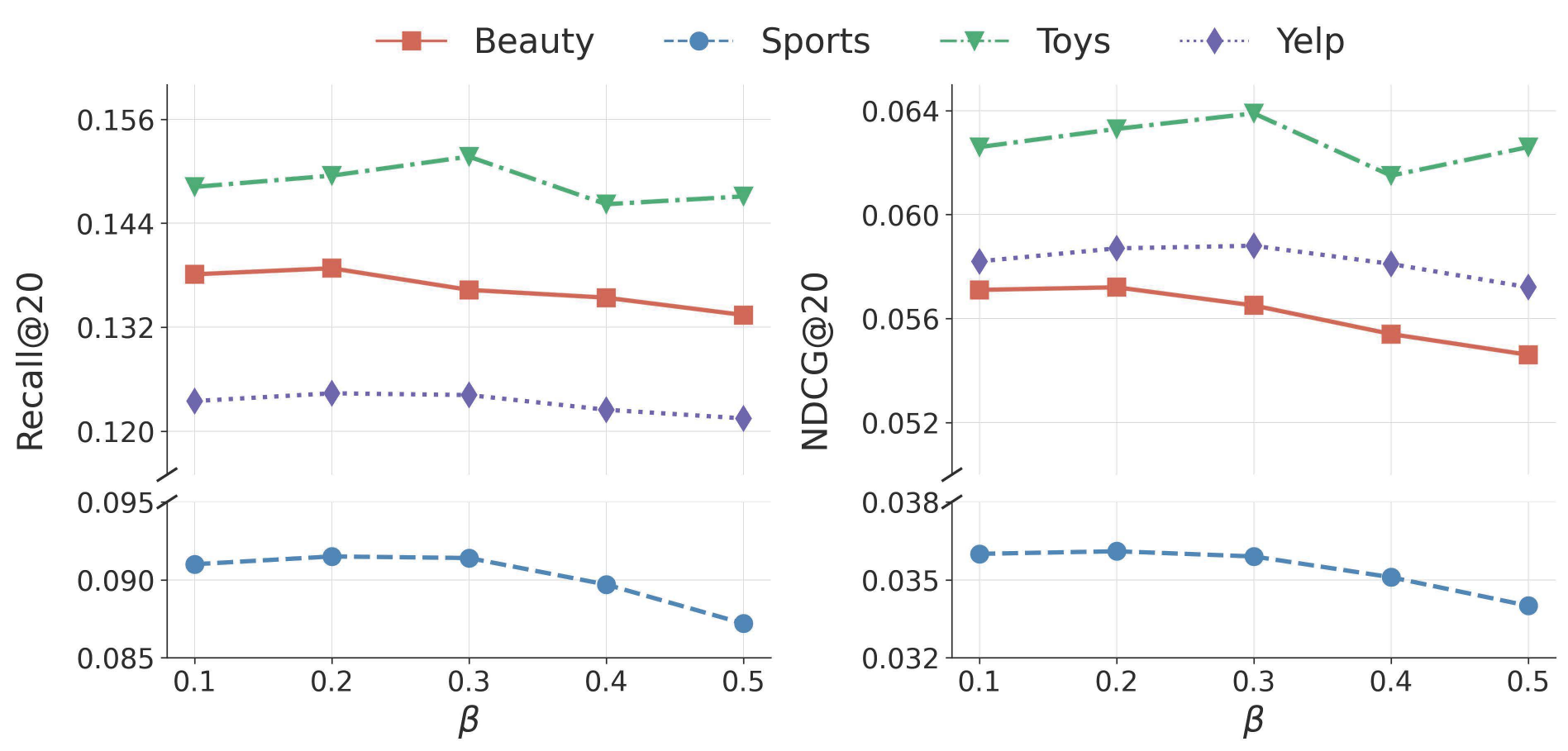}
   \caption{Effects of attribute prediction weights $\beta$.}
   \Description{}
   \label{fig:7}
\end{figure}

\vspace{1mm}
\noindent\textbf{Attribute Prediction Weights}. We then analyzed the effect of the attribute prediction weight, $\beta$. As depicted in Fig.~\ref{fig:7}, the optimal $\beta$ value varies across datasets. The Toys datasets reach their peak performance with a $\beta$ weight of 0.3, while the Yelp and Sports dataset performs optimally and similarly within the 0.2 to 0.3 weight range. In contrast to these, the Beauty dataset achieves its best results with a weight of 0.1. This result highlights the varying importance of category attribute information for enhancing recommendation performance across different domains.

\subsection{Training Efficiency}\label{Sec: efficiency}
\begin{table}[!t]
    \centering
    \caption{Comparison of training cost. \#Param: number of tunable parameters;  Time/E: average training time for one epoch, measured in seconds (`s'); MU: GPU memory usage.}
    \label{tab8:efficience}
    \resizebox{0.48\textwidth}{!}{
        \begin{tabular}{c|c|c|c|c|c|c}
            \toprule
            Datasets & Model & Recall@20 & NDCG@20 & \#Param & Time/E & MU \\        
            \midrule
            \multirow{3}{*}{Yelp} 
            & MSSR & 0.1003 & 0.0495 & 8.7M & 52s & 6.5G \\
            & DIFF & \underline{0.1164} & \underline{0.0554} & 7.1M & 33s & 8.1G \\
            & TASIF & \textbf{0.1254} & \textbf{0.0589} & 7.2M & 28s & 5.0G \\
            \midrule
            \multirow{3}{*}{Beauty} 
            & MSSR & \underline{0.1274} & 0.0535 & 6.0M & 26s & 4.3G \\
            & DIFF & 0.1216 & \textbf{0.0587} & 5.2M & 16s & 4.5G \\
            & TASIF & \textbf{0.1381} & \underline{0.0564} & 4.8M & 10s & 2.7G \\
            \midrule
            \multirow{3}{*}{Sports} 
            & MSSR & 0.0799 & 0.0321 & 8.4M& 42s & 6.5G \\
            & DIFF & \underline{0.0804} & \textbf{0.0358} & 6.8M & 27s & 8.1G \\
            & TASIF & \textbf{0.0862} & \underline{0.0353} & 6.8M & 23s & 6.0G \\
            \midrule
            \multirow{3}{*}{Toys} 
            & MSSR & \underline{0.1370} & 0.0590 & 6.2M & 24s & 5.6G \\
            & DIFF & 0.1323 & \underline{0.0607} & 5.1M & 15s & 7.3G \\
            & TASIF & \textbf{0.1471} & \textbf{0.0617} & 4.9M & 11s & 3.9G \\
            \bottomrule
        \end{tabular}
    }
\end{table}

To validate the efficiency and effectiveness of our proposed TASIF, we conducted comprehensive comparisons against the baseline MSSR and the state-of-the-art method DIFF. 
For a fair comparison, all models were configured with two Transformer layers, two attention heads, and a batch size of 512. Other hyperparameters were set to the optimal values reported in their respective papers.

As detailed in Table~\ref{tab8:efficience}, the results clearly demonstrate TASIF's performance. In terms of recommendation effectiveness, TASIF achieves the highest Recall@20 across all four datasets and delivers highly competitive NDCG@20 scores, underscoring its recommendation quality. Crucially, this high quality does not come at the cost of efficiency. On the contrary, TASIF consistently operates with fewer parameters, requires the shortest training time per epoch, and consumes the least GPU memory. These findings provide robust evidence that TASIF strikes a desirable balance, advancing recommendation effectiveness while establishing new benchmarks for training cost and efficiency.

\subsection{Case Study}
\begin{figure}[h]
  \centering
   \includegraphics[width=\linewidth]{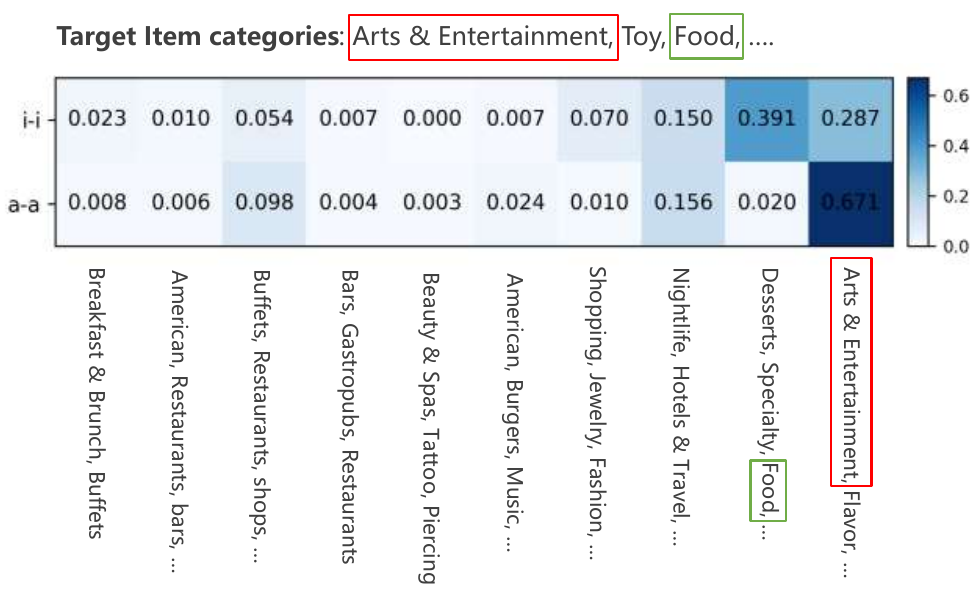}
   \caption{A case study on attention distributions within item ID sequences (`i-i') and attribute sequences (`a-a') based on an actual user. The horizontal axis shows the categories of the items in the user sequence. The target categories refer to the ground truth categories of the next item to be predicted.}
   \Description{}
   \label{fig:8}
\end{figure}

Finally, we conducted a case study on the Yelp dataset using an actual user sequence, as shown in Fig.~\ref{fig:8}. 
We observe that the attribute representation attends strongly to the target category ``Arts \& Entertainment'', which suggests that the model may have captured user-relevant attribute patterns. Moreover, the item ID representation, enhanced through fusion with attribute information, focuses on another target category ``Food'', further implying the effectiveness of our fusion mechanism in highlighting semantically meaningful signals.

\section{CONCLUSIONS}
In this paper, we proposed TASIF, a novel framework that overcomes inefficient side information fusion and inadequate temporal modeling in sequential recommendation. The effectiveness of TASIF stems from three synergistic components: a Time Span Partitioning (TSP) module for global time awareness, an Adaptive Frequency Filter (AFF) for flexible denoising, and a highly efficient Adaptive Side Information Fusion (ASIF) layer. Extensive experiments demonstrate that TASIF achieves state-of-the-art performance while simultaneously enhancing computational efficiency and temporal modeling, confirming the efficacy of our design.

\section*{ETHICAL CONSIDERATIONS}
Our research was conducted in strict adherence to ethical guidelines. We utilized publicly available benchmark datasets for all our experiments. Any user-related information within these datasets had been fully anonymized by the data providers, ensuring the protection of user privacy. We have complied with the data usage policies for each dataset, employing them solely for the academic purpose of validating our proposed model's effectiveness, with no commercial interests involved. Furthermore, the model we propose is designed for recommendation tasks and does not inherently pose any security or safety risks to end-users.

\bibliographystyle{ACM-Reference-Format}
\bibliography{content/references}

\appendix

\end{document}